\documentclass[preprint,12pt]{elsarticle}

\usepackage[version=3]{mhchem} 
\usepackage{siunitx}
\usepackage{graphicx}
\usepackage{amssymb}
\usepackage{amsmath}
\usepackage{booktabs}
\usepackage{multirow}
\usepackage{tabularx}
\usepackage{ulem}
\usepackage{mciteplus}
\usepackage{caption, subcaption}
\usepackage{afterpage}
\usepackage{float}
\usepackage{lineno}
\usepackage{lastpage}
\usepackage{fancyhdr}
\usepackage[T1]{fontenc}
\usepackage[utf8]{inputenc}

\usepackage{amssymb}
\usepackage{amsmath}

\journal{Journal of Colloid and Interface Science}

\begin{document}

\begin{frontmatter}




\title{Interfacial Behavior from the Atomic Blueprint: Machine Learning-Guided Design of Spatially Functionalized $\alpha$-\ce{SiO2} Surfaces}


\author[lemta]{Evgenii Strugovshchikov}
\ead{evgenii.strugovshchikov@univ-lorraine.fr}

\author[lemta]{Viktor Mandrolko}

\author[lpct]{Dominika Lesnicki}

\author[lpct]{Mariachiara Pastore}

\author[lemta]{Laurent Chaput}

\author[lemta]{Mykola Isaiev}

\affiliation[lemta]{
  organization={Université de Lorraine, CNRS, LEMTA},
  addressline={},
  city={Nancy},
  postcode={F-54000},
  state={},
  country={France}
}

\affiliation[lpct]{
  organization={Université de Lorraine \& CNRS, LPCT, UMR 7019},
  addressline={},
  city={Nancy},
  postcode={F-54000},
  state={},
  country={France}
}

\begin{abstract}
$\alpha$-Quartz (\ce{SiO2}) surfaces functionalized with hydroxyl (\ce{OH}) and methyl (\ce{CH3}) groups provide a versatile platform for controlling interfacial properties critical to applications such as catalysis, protective coatings, and energy conversion. 
The arrangement of these functional groups strongly influences interfacial interactions at solid–liquid interfaces, highlighting their relevance to colloid and interface science.
However, conventional models often treat surface functionalization as spatially homogeneous, overlooking the atomic-scale organization of surface groups. 
We hypothesize that this spatial distribution, beyond overall composition, plays a decisive role in governing surface stability and interfacial behavior.

To test this hypothesis, we employ a multi-scale simulation workflow combining density functional theory (DFT), ab initio molecular dynamics (AIMD), and machine-learned force fields (MLFFs).
This approach allows us to explore a range of spatial patterns of \ce{OH}/\ce{CH3} functionalization on the $\alpha$-quartz (0001) surface.
We evaluate the impact of spatial arrangements on mixing energy, hydrogen bonding networks, and vibrational properties with high accuracy and robustness.

Our results reveal that spatial patterning strongly influences surface stability and interfacial structure. 
A thermodynamically favored unpaired configuration emerges near 67\% \ce{CH3} substitution, where isolated \ce{OH} groups form secondary hydrogen bonds through reorientation toward subsurface oxygen atoms. 
This rearrangement induces a characteristic blue shift in \ce{OH} stretching frequencies, indicating weaker H-bonding. 
These effects are absent in clustered arrangements. 
By establishing a clear link between functional group patterning and interfacial behavior, our work uncovers the underlying mechanisms to guide and accelerate the rational design of silica-based materials and coatings, directly relevant to colloid and interface science. 
\end{abstract}

\begin{graphicalabstract}
\centering
\includegraphics[width=0.95\textwidth]{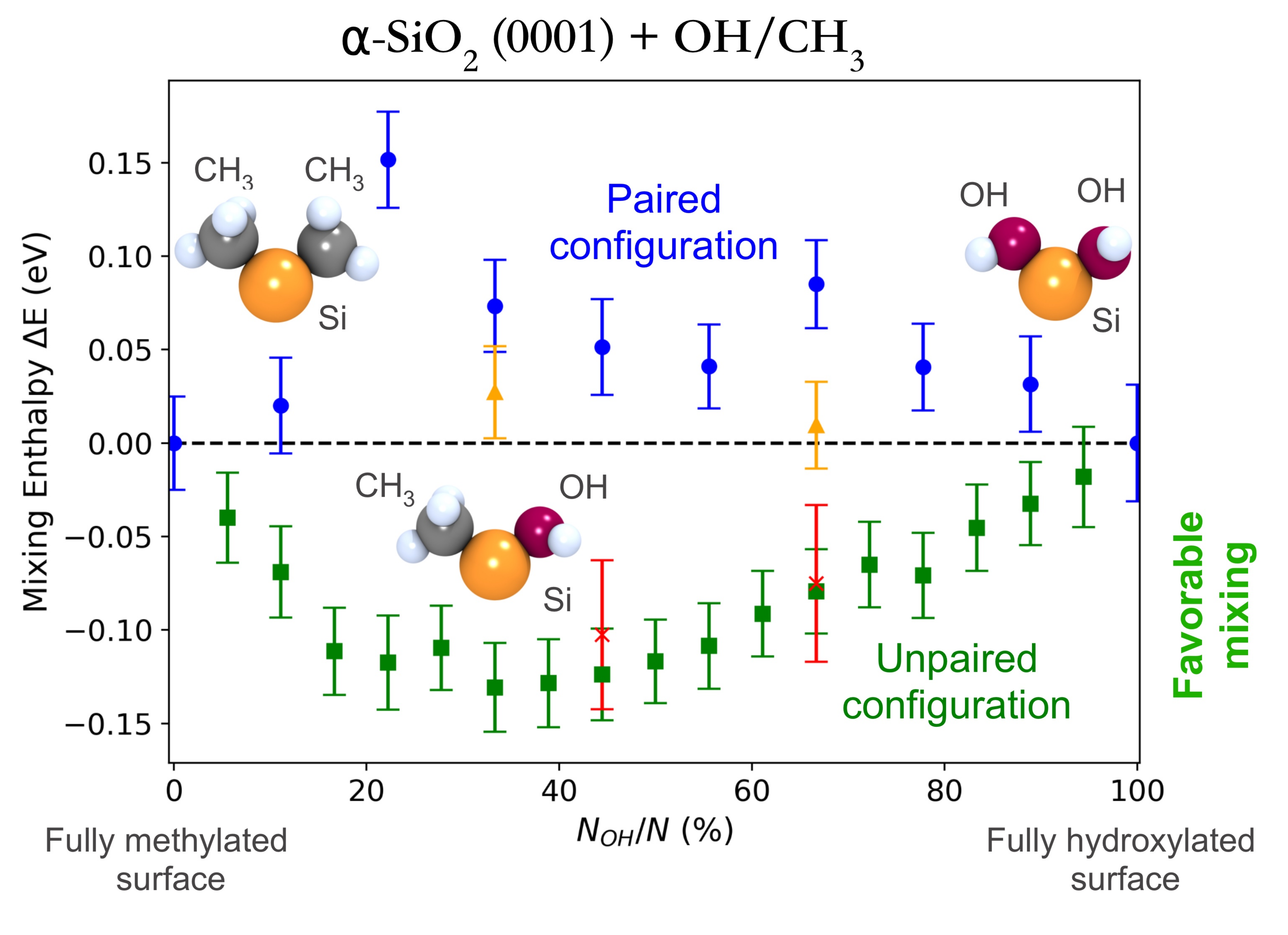}
\end{graphicalabstract}

\begin{highlights}
\item On-the-fly machine-learned force fields (MLFFs) enable accurate, large-scale modeling of chemically diverse surfaces.
\item Spatial patterning of hydroxyl and methyl groups controls surface stability and interface behavior.
\item Isolated hydroxyl groups are stabilized by forming secondary hydrogen bonds in mixed functionalization patterns.
\end{highlights}


\end{frontmatter}




\section{Introduction}
\label{sec1}

Functionalization of oxide surfaces, particularly silicon dioxide in its $\alpha$-quartz ($\alpha$-\ce{SiO2}) form, has emerged as a powerful strategy for tailoring material behavior in critical technologies, including catalysis, sensing, energy conversion, and protective coatings\cite{Ye2023,Wang2021,Liu2016,Mercado2022}. 
By chemically modifying the surface with hydroxyl (\ce{OH}) and methyl (\ce{CH3}), it is possible to finely tune interfacial properties such as hydrophilicity, adsorption dynamics, and chemical reactivity\cite{Cheng2020,Luechinger2005,Aono2016,Kodama2021}.
Many practical silanization and surface-coating processes result in chemically heterogeneous terminations, where hydroxyl (\ce{OH}) and methyl (\ce{CH3}) groups coexist rather than forming fully hydroxylated or fully methylated surfaces. 
For example, functionalization with alkylsiloxane reagents often leads to incomplete substitution, leaving behind residual silanol sites and producing mixed hydrophilic–hydrophobic domains on the surface \cite{Vukovic2023}. 
Experimental studies have shown that this type of chemical heterogeneity can significantly influence macroscopic interfacial properties, such as wettability and contact angle behavior, compared to uniformly functionalized surfaces \cite{Vukovic2023, Warring2016}. 
Complementary theoretical investigations on silica polymorphs, such as tridymite and glass, further demonstrate that partial functionalization yields water–surface interactions and wetting responses that differ markedly from those of single-terminated models \cite{Wang2022}.

While prior studies have predominantly focused on the overall concentration of these functional groups, growing evidence reveals that their spatial distribution at the atomic scale is a key determinant of surface behavior\cite{Yang2009}.
Subtle variations in the local arrangement of \ce{OH} and \ce{CH3} groups can dramatically alter hydrogen-bond networks, surface energetics, and ultimately macroscopic phenomena such as wettability and stability.
Despite its profound implications, this level of structural control remains poorly understood and underexplored\cite{Gartmann2011}.
Closing this gap is essential for advancing the rational design of next-generation materials, where nanoscale chemical patterning could unlock unprecedented control over interfacial functionality.

Various experimental techniques are utilized to analyze the structure of such surfaces. 
While direct visualization of atomic-scale arrangements remains difficult, a combination of advanced surface-sensitive methods has enabled partial spatial and orientational characterization of surface functional groups. 
Vibrational techniques such as FTIR and sum-frequency generation (SFG) spectroscopy can distinguish free versus hydrogen-bonded \ce{OH} groups and probe their orientation relative to the surface normal\cite{Rimola2013,Ye2001,Pezzotti2019}. 
Solid-state NMR, particularly \ce{^{29}Si} and \ce{^{1}H} MAS, offers complementary insights into silanol speciation and \ce{Si}–\ce{C} bonding environments\cite{Murray2010,Zhao2012}. 
Additionally, ToF-SIMS and atomic force microscopy (AFM) have been employed to spatially resolve \ce{OH} and \ce{CH3} domains at micro- to nanoscale resolution\cite{Wang2014,Daza2019}. 
Together, these methods provide indirect yet powerful tools for assessing spatial heterogeneity and chemical functionality on $\alpha$-quartz surfaces.
Despite these advances, a complete atomistic picture of surface functionalization remains elusive. 
In this context, computational methods have become essential for gaining detailed insights into surface phenomena\cite{Pezzotti2021,Khanniche2017,Khanniche2012}. 
Density functional theory (DFT) and molecular dynamics (MD) simulations have proven particularly effective in capturing both the electronic structure and the dynamic behavior of functionalized surfaces\cite{Segalina2025,Segalina2021,Foucaud2023,Potts2021,Sundararaman2022,Uhlig2021}.

In the paper by Chen et al.\cite{Chen2011}, \textit{ab initio} simulations were employed to investigate the functionalization of $\alpha$-\ce{SiO2} surfaces. 
One study specifically focused on the $\alpha$-\ce{SiO2} ($0001$) surface functionalized with hydroxyl (\ce{OH}) groups, analyzing their structural arrangement and interactions with water molecules. 
The results provided valuable insights into the formation of a hydrogen-bond network among surface \ce{OH} groups, emphasizing its role in surface reactivity and wettability.
Molecular dynamics (MD) simulations have also been extensively used to study functionalized $\alpha$-\ce{SiO2} surfaces, particularly their behavior in aqueous environments. 
A notable advancement in this field was the development of the CLAYFF force field\cite{Cygan2004}, designed for the simulation of hydrated and multicomponent mineral systems and their interfaces with water. 
The force field parameters were derived from density functional theory (DFT) calculations of simple oxides, hydroxides, and oxyhydroxides, ensuring accurate representation of surface interactions in complex environments.
Building on this, CLAYFF was later adapted to model functionalized $\alpha$-\ce{SiO2} surfaces, incorporating additional force fields such as DREIDING to account for organic functional groups (-\ce{CH_x}). 
The newly developed potential, refined with additional DFT-based training, enabled investigation of the wettability of hydroxylated $\alpha$-\ce{SiO2} surfaces while accounting for the effects of co-adsorbed pentyl and hydroxyl groups\cite{Abramov2019}.
Further advancements in molecular modeling led to the development of realistic models for hydroxylated, ethoxylated, and methylated silica surfaces. 
It was noted that the molecular modeling of methylated silica surfaces remained limited, largely relying on the framework established by Furukawa et al\cite{Furukawa2005}. 
Their approach employed a united-atom model to represent silylated or trimethylated (-\ce{Si(CH3)3}) silica surfaces, treating methyl groups as single interaction sites. 
More recent studies extended this work to different $\alpha$-\ce{SiO2} surface orientations\cite{Khanniche2017}; however, their parameterization was restricted to surfaces functionalized with \ce{Si(CH3)2} double groups, leaving a gap in the understanding of more diverse functionalized surface chemistries.

Addressing this gap requires atomistic simulations that can capture large, compositionally complex systems with sufficient accuracy, yet developing such models for previously unexplored chemical configurations remains computationally demanding and often requires manual, system-specific parameterization \cite{Wu2023}.
Recent advances in machine learning force fields (MLFFs) have offered a promising solution to these challenges\cite{Jacobs2025, Thiemann2024, Cheng2020-2}. 
Traditional modeling approaches struggle with the trade-off between accuracy and system size.
The advent of machine learning force fields (MLFFs) represents a methodological leap, allowing atomistic accuracy to scale across realistic systems\cite{Wu2023}. 
Their adaptive nature enables dynamic exploration of complex chemistries, making them especially suited for mapping the energetics of heterogeneous, functionalized surfaces\cite{Bag2021}.
By integrating machine learning with on-the-fly active learning schemes, MLFFs can automatically generate accurate interatomic potentials during simulations without the need for extensive pre-training\cite{Liu2024}. 
This methodology has been successfully applied to complex systems such as liquid/solid interfaces\cite{Li2023,Chen2025,Romano2025,Liu2024}, inorganic-organic hybrid materials\cite{GarroteMrquez2024,Wieser2024} and macroscopic materials properties\cite{Jinnouchi2019,Legenstein2025}, where it enabled large-scale simulations with accuracy comparable to that of \textit{ab initio} MD. 

In this study, we leverage a combination of DFT, \textit{ab initio} MD, and on-the-fly MLFF techniques to systematically investigate the effect of the \ce{OH}/\ce{CH3} spatial distribution on the surface properties of $\alpha$-\ce{SiO2}.
Focusing on the analysis of the hydrogen bonding network, vibrational spectra, and energetics, we aim to investigate how different spatial arrangements, ranging from uniformly dispersed to clustered, affect both the stability and the macroscopic behavior of the surface.
Our results highlight a distinct regime of \ce{OH} functionalization, wherein the mixing energy reaches a minimum and correlates with a marked change in surface properties, underscoring the importance of local chemical environments.
Thus, our study introduces a new conceptual framework where the spatial distribution, rather than just the concentration, of functional groups acts as the primary descriptor of surface behavior. 
This shift in perspective offers a richer understanding of surface reactivity and wettability, potentially redefining how functionalization strategies are approached in materials design.

By bridging the gap between atomic-scale arrangements and macroscopic properties, our work not only advances the fundamental understanding of functionalized surfaces but also lays the groundwork for the rational design of next-generation materials. 
These findings have indeed direct implications for the design of high-performance coatings, selective adsorption layers in environmental remediation, and stable, low-energy surfaces in microelectronic and photonic devices, where tailored wettability and surface energy are critical parameters.

\section{Methodology}
\label{sec2}

\subsection{On-the-fly training of ML Force Field}  
\label{subsec1}

\textit{Ab initio} molecular dynamics (AIMD) simulations were performed to generate training data for on-the-fly machine learning force field development.
The simulations were carried out using the Vienna Ab initio Simulation Package (VASP 6.4.3)\cite{Kresse1996} with the PBE-GGA functional and projector augmented wave (PAW) pseudopotentials.\cite{Perdew1996,Kresse1999} 
The plane-wave basis set included $3s^{2}3p^{2}$, $2s^{2}2p^{4}$, $2s^{2}2p^{2}$, and $1s^1$ valence electrons for \ce{Si}, \ce{O}, \ce{C} and \ce{H} atoms, respectively. 
Dispersion interactions were accounted for using Grimme’s DFT-D3 correction.\cite{Grimme2010}
A plane-wave energy cutoff of $500$ eV was used. 
Due to computational constraints, Brillouin zone sampling was restricted to the $\Gamma$-point within the Monkhorst-Pack framework~\cite{Monkhorst1976}.
The convergence criteria were set to $10^{-3}$  eV for the electronic minimization.

In order to gain insight into the energetic stability and possible preferential spatial arrangements of surface functional groups, we analyze the mixing energy, which quantifies the thermodynamic favorability of combining hydroxyl (\ce{OH}) and methyl (\ce{CH3}) groups on the same $\alpha$-\ce{SiO2} (0001) surface. 
Specifically, the mixing energy measures deviations from ideal behavior, where ideal mixing would imply no energetic preference for segregation or clustering of functional groups:

\begin{equation}
E_{\text{mix}} = E_{\text{surf+\ce{OH}/\ce{CH3}}} - x E_{\text{surf+\ce{OH}}} - (1-x) E_{\text{surf+\ce{CH3}}}
\end{equation}

Here, $E_{\text{surf+\ce{OH}/\ce{CH3}}}$ denotes the total energy of the surface with a mixed distribution of \ce{OH} and \ce{CH3} groups, while $E_{\text{surf+\ce{OH}}}$ and $E_{\text{surf+\ce{CH3}}}$ correspond to the total energies of surfaces fully terminated with \ce{OH} and \ce{CH3}, respectively. 
The parameter $x$ represents the fraction of hydroxyl groups on the surface. 
A negative mixing energy indicates a thermodynamically favorable interaction between the two species, potentially promoting homogeneous mixing, while a positive value suggests a tendency toward phase separation or surface patterning.

All AIMD simulations were conducted in the canonical ensemble (NVT) using a Langevin thermostat to maintain a constant temperature of $300$ K.  
The Velocity-Verlet integration algorithm was employed to propagate atomic trajectories, ensuring accurate integration of the equations of motion.  
A time step of $0.5$ fs was used to maintain numerical stability and precise energy conservation, particularly due to the presence of light hydrogen atoms in the system.  
The total simulation time for each training MD run was set to $10$ ps ($20000$ steps).  

The machine learning force field was trained iteratively by accumulating reference data from AIMD simulations at successive stages of surface functionalization.
The training set consists of energies, forces and stresses calculated using DFT simulations.
During active learning, a machine-learned force field is initially trained on ab initio data collected from the first few timesteps of AIMD simulations. 
The trajectory then evolves using the MLFF, provided that the uncertainties in the predicted forces, estimated via Bayesian errors, remain below a predefined threshold\cite{Jinnouchi2019}. 
If the uncertainty exceeds this threshold, the MLFF prediction is discarded, and the next timestep is instead computed using DFT. 
The resulting DFT-calculated configuration is subsequently incorporated into the training set, enabling an on-the-fly refinement of the MLFF. 
This iterative approach ensures that only a few hundred of the most relevant configurations are selected for DFT calculations, despite the trajectory spanning several thousand timesteps.


\subsection{Training dataset}
\label{subsec2}

To develop the training set, first we performed DFT optimizations of $\alpha$-\ce{SiO2} (0001) surface functionalized with varying \ce{OH}/\ce{CH3} ratios.
Details of the static DFT calculations are provided in the Supporting Information. 
The DFT-calculated mixing energies of the functionalized surfaces are shown in Figure~S1 of the Supporting Information. 
The interaction energy between the functional groups were also analyzed, with the results presented in Figure~S2 of the Supporting Information. 
These calculations were carried out to identify the most stable configurations corresponding to different spatial distributions of the functional groups.

\begin{figure}[!htb]
\centering
     \includegraphics[width=0.95\textwidth]{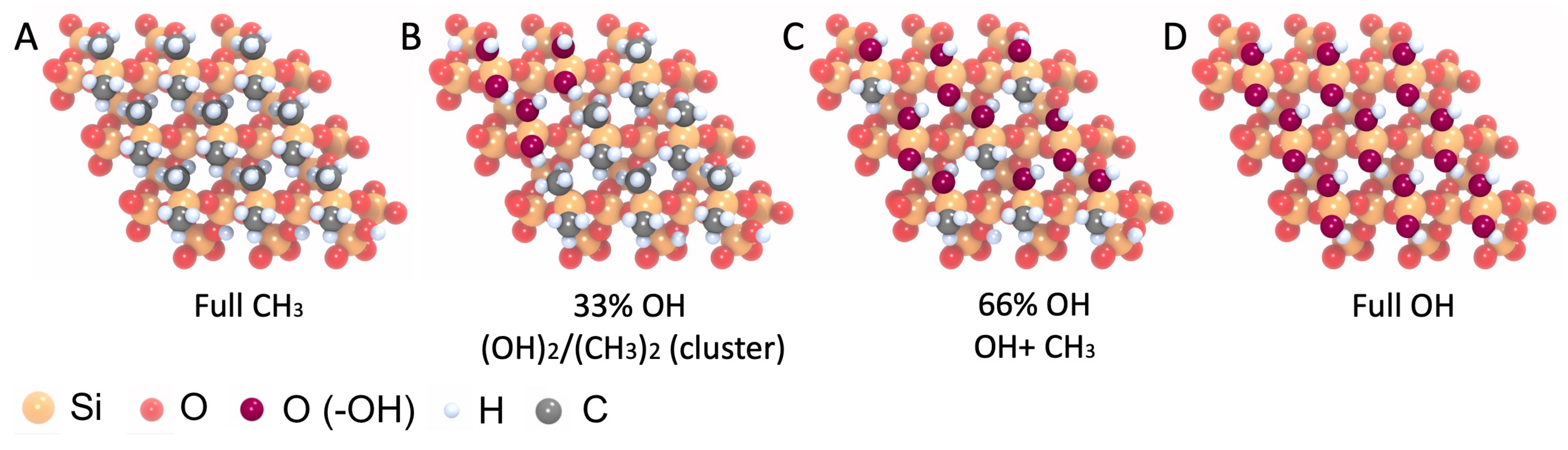}
\caption{\label{fig:dft_structures} 
Snapshots of the mixed functionalization of the $\alpha$-\ce{SiO2} (0001) surface with varying \ce{OH}/\ce{CH3} ratios. The DFT-optimized structures illustrate the spatial distribution of functional groups across different compositions. The silicon atoms are shown in yellow, framework oxygen atoms (\ce{SiO2}) in orange, hydroxyl oxygen atoms in red, hydrogen atoms in white, and carbon atoms in grey.
}
\end{figure}

The on-the-fly AIMD training process was performed iteratively, beginning with bulk $\alpha$-\ce{SiO2} and progressively extending to functionalized surfaces.  
Figure~\ref{fig:dft_structures} presents DFT-optimized snapshots of the $\alpha$-\ce{SiO2} (0001) surface with mixed functionalization at varying \ce{OH}/\ce{CH3} ratios, illustrating the spatial distribution of functional groups.
For clarity, we refer to configurations in which both functional groups on a single surface silicon atom are the same, either \ce{Si-(OH)2} or \ce{Si-(CH3)2}, as \textit{paired} configurations. 
In contrast, when a single \ce{Si} atom is bonded to one hydroxyl and one methyl group (\ce{OH-Si-CH3}), we define this as an \textit{unpaired} configuration.
Accordingly, if a surface structure contains at least one unpaired configuration, it is classified as an \textit{unpaired structure}; otherwise, it is considered a \textit{paired structure}.

By analyzing the mixing energies, the most energetically favorable arrangements of \ce{OH} and \ce{CH3} groups were determined. 
Based on DFT insights, we initiated the training of functionalized $\alpha$-\ce{SiO2} surfaces by fully terminating them with methyl (\ce{Si-(CH3)2}) and hydroxyl (\ce{Si-(OH)2}) groups to capture their respective local environments (shown in Figure~\ref{fig:dft_structures}A and~\ref{fig:dft_structures}D).
Subsequently, mixed \ce{OH}/\ce{CH3} configurations were incorporated to account for their interactions and to extend the training set to more complex surface chemistries. 

As a case study, three characteristic \ce{OH}/\ce{CH3} configurations were selected:  
\begin{enumerate}
    \item An unpaired \ce{OH-Si-CH3} configurations with a relative concentration of \ce{OH} groups of $67$ \%, to explore the effect of unpaired distribution (Figure~\ref{fig:dft_structures}C).  
    \item A paired \ce{Si-(OH)2}/\ce{Si-(CH3)2} cluster configuration at the same \ce{OH} concentration of $67$ \%, to evaluate how different local environments influence the energetic properties of the system.  
    \item A paired \ce{Si-(OH)2}/\ce{Si-(CH3)2} structure with a  \ce{OH} concentration of $33$ \%, to explore the effects of a different functional group ratio (Figure~\ref{fig:dft_structures}B).  
\end{enumerate}

Throughout the training process, the root mean square errors (RMSE) for energy and forces, extracted from AIMD calculations, were used to iteratively refine the force field. 
This approach enabled the model to systematically learn interactions, starting from bulk \ce{SiO2} and progressing toward increasingly complex surface chemistries.  
During all the simulations, the maximum RMSE values did not exceed $1.0$ meV/atom for energy and $75$ meV/\AA ~ for forces, ensuring the reliability of the trained force field.
The resulting training set contains $2330$ DFT snapshots, from which $2100$ local reference configurations for \ce{Si}, \ce{O}, \ce{H} and \ce{C} are extracted.

After training all configurations, a refitting process was performed. 
As a result of extensive hyperparameter search\cite{Jinnouchi2023}, the cutoff radii for both angular and radial descriptors were adjusted to $13$ \AA ~and $5$ \AA, respectively. 
Following hyperparameter optimization of the resulting force field across the entire training set, the final errors for energy and forces were $0.67$ meV/atom and $63$ meV/\AA, respectively.

\subsection{Accuracy and error analysis}  
\label{subsec3}

\begin{figure}[!htb]
\centering
     \includegraphics[width=0.75\textwidth]{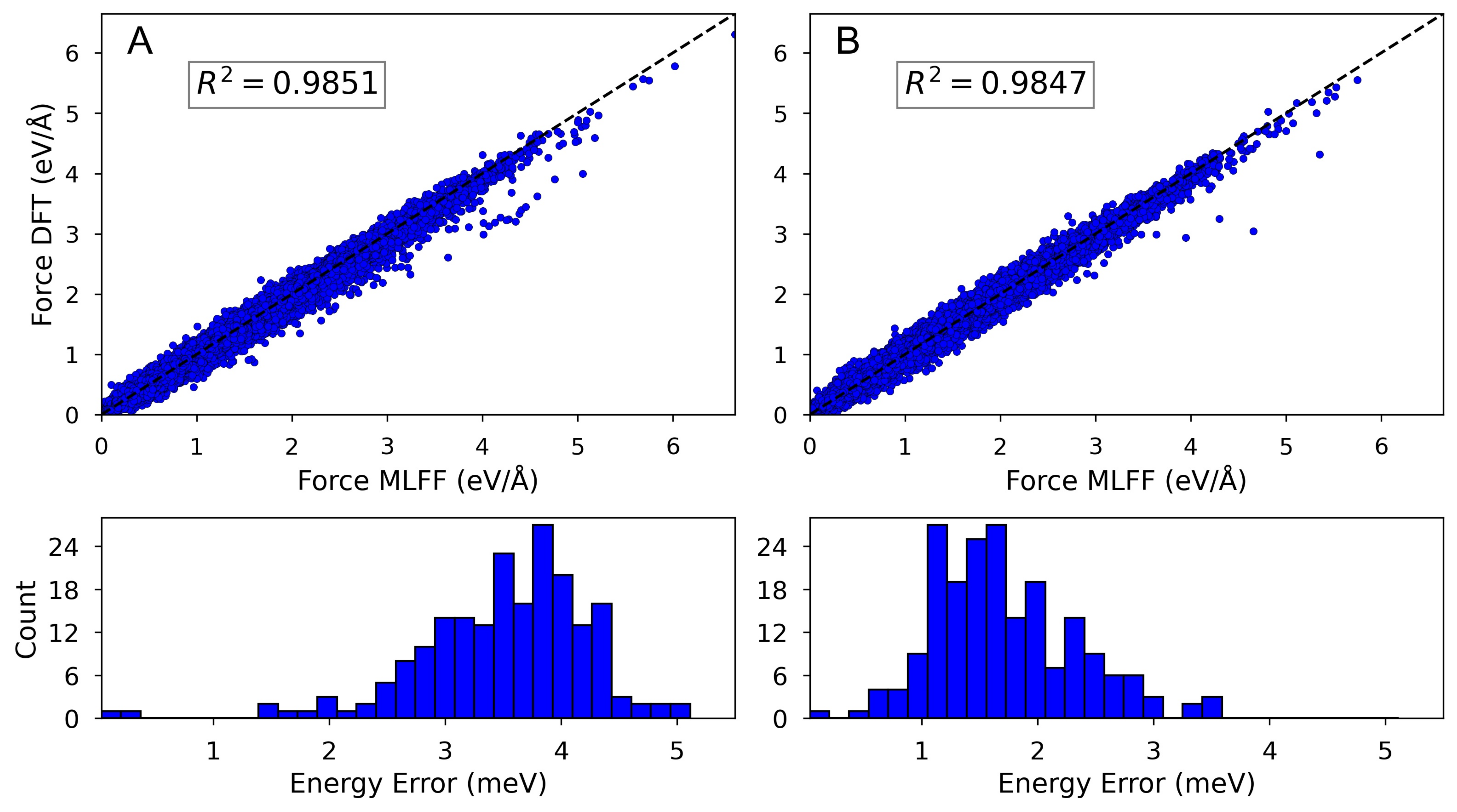}
\caption{\label{fig:forces_validation} 
The external validation of the ML potential by assessing the forces and energy errors calculated by MLFF and verified by means of single point DFT calculation. }
\end{figure}

The developed machine-learned force field for the functionalized $\alpha$-\ce{SiO2} surface was validated against DFT calculations by assessing its ability to predict system behavior beyond the training set. 
Figure~\ref{fig:forces_validation} presents a comparison of the forces and energy error distribution from the validation set as predicted by both DFT and the generated MLFF potential.
As a validation set, we used functionalized surfaces that were not included in the training process, specifically unpaired \ce{OH-Si-CH3} configurations with relative \ce{OH} concentrations of $17\%$ (Figure~\ref{fig:forces_validation}A) and $44\%$ (Figure~\ref{fig:forces_validation}B). 
First, these structures were used to perform MD simulations of $5$ ps using the trained MLFF. 
Subsequently, $200$ configurations were randomly selected from the resulting trajectories and validated using DFT with the same parameters as those used for training.  
The root mean square errors (RMSE) for energy and forces were evaluated as $2.7$ meV/atom and $89$ meV/\AA, respectively, demonstrating the accuracy of the developed force field in capturing the interactions of mixed functional groups on the $\alpha$-\ce{SiO2} surface.  

As in the previous studies\cite{Jinnouchi2019,MonterodeHijes2024,Jinnouchi2024}, the validation RMSE values fall within the predefined acceptable range for MLFFs trained using semi-local functionals with dispersion corrections, specifically, $1-5$ meV/atom for energy and $100$ meV/\AA~for forces.
The distribution of energy differences between the DFT and MLFF predictions follows a Gaussian profile, with a mean value centered around mean absolute errors. 
These results indicate that the force field is well-fitted and capable of accurately representing a diverse range of structural and chemical environments.

\section{Results and discussions}
\label{sec3}

\subsection{Spatial distribution of the functional groups}
\label{subsec4}

The newly parameterized MLFF was first applied in MD simulations of functionalized $\alpha$-\ce{SiO2}  (0001) surfaces to investigate the spatial distribution of functional groups at a finite temperature of $300$ K. 
To systematically explore a broad range of functionalization patterns, the simulations started with a surface fully covered by \ce{OH} groups, followed by a stepwise substitution of \ce{OH} with \ce{CH3} groups.  
This approach mirrors experimental synthesis procedures, where hydroxyl-terminated surfaces are typically prepared as an initial step.
At each substitution level (fixed \ce{OH}/\ce{CH3} ratio), we generated and simulated a variety of paired and unpaired arrangements, clustered, non-clustered, and uniformly dispersed, to identify the most energetically stable configurations.

\begin{figure}[!htb]
\centering
     \includegraphics[width=0.9\textwidth]{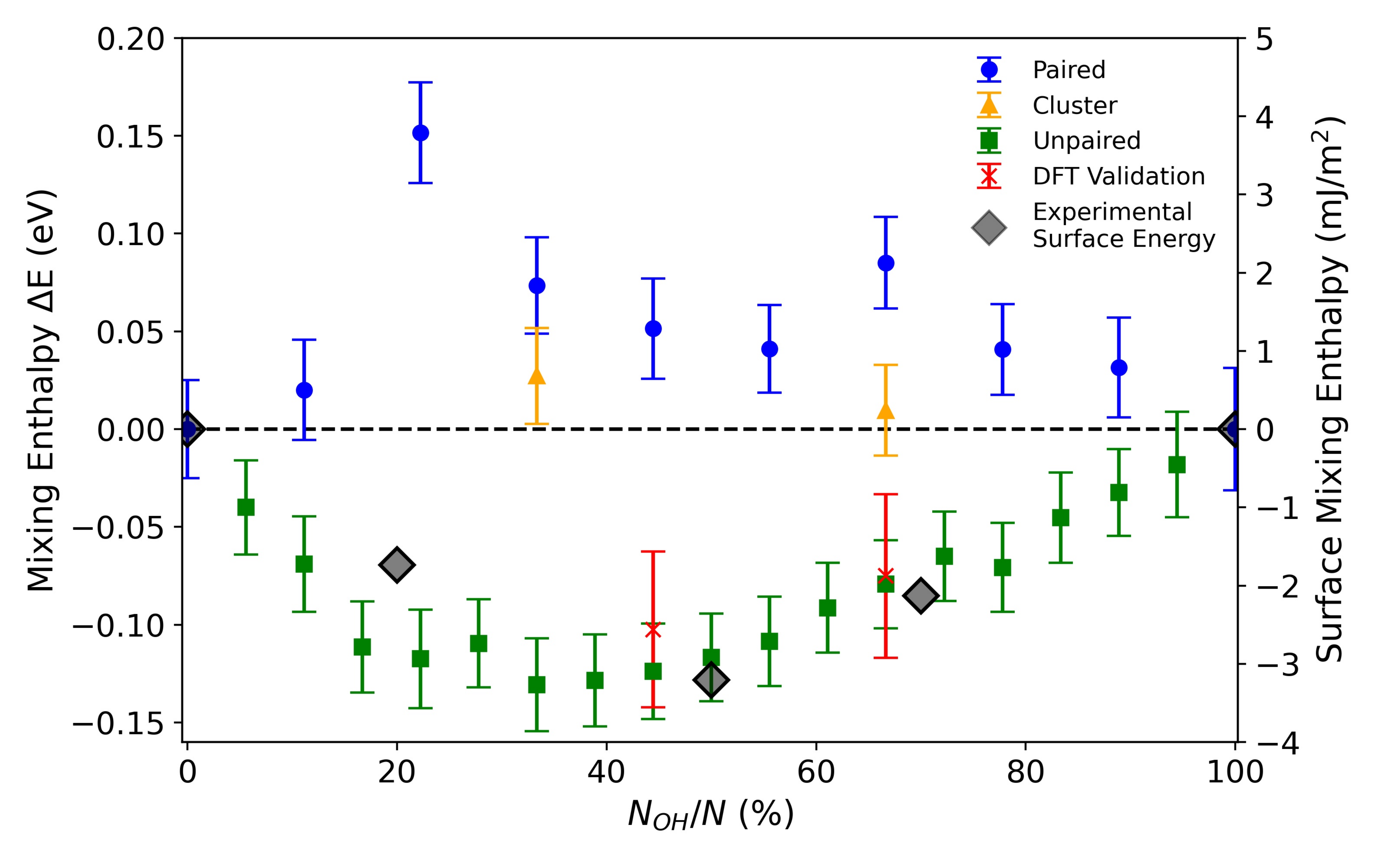}
\caption{\label{fig:mixing_energy} 
Mixing enthalpy of \ce{OH}/\ce{CH3} functional group distributions on the $\alpha$-\ce{SiO2} (0001) surface, calculated using the average total free energy from MD simulations at $300$ K. Each data point represents the average over $2$ independent surface models, with statistics collected from $80000$ MD steps. Green represents unpaired \ce{OH-Si-CH3} configurations, while blue and yellow indicate paired \ce{Si-(OH)2}/\ce{Si-(CH3)2} spread and clustered configurations, respectively. Red denotes DFT validation points. The mixing enthalpy is normalized by the number of surface \ce{Si} atoms. Black rhombs correspond to experimental surface mixing enthalpies extracted from the work of Yang et al.~\cite{Yang2009}}
\end{figure}

After an initial equilibration phase of $10$ ps, production MD simulations were conducted for an additional $40$ ps. 
The total free energy of the system, including both potential and kinetic contributions, was averaged over all simulation steps to analyze the energetic trends associated with different functional group distributions at $300$ K.
The resulting values were plotted as a function of the \ce{OH}/(\ce{OH}+\ce{CH3}) ratio, as shown in Figure~\ref{fig:mixing_energy}. 
The selected data points in the graph, not included in the MLFF training, were validated using ab initio MD simulations at the corresponding \ce{OH}/\ce{CH3} ratios and finite temperature (marked in red). 
The results showed that the mixing enthalpies obtained from MLFF closely match those from DFT-based simulations, confirming the accuracy of the MLFF in predicting the distribution of functional groups.
 
The energy distribution obtained from MD simulations closely follows the trends observed in DFT-optimized calculations (Figure~S1 of SI), indicating that temperature effects at $300$ K do not significantly alter the fundamental characteristics of functional group distribution. 
An energy minimum is observed at a \ce{CH3} substitution ratio of $67\%$ ($33\%$ of \ce{OH}), indicating this configuration as the most stable one.  
This minimum appears exclusively for the unpaired distribution of functional groups, where surface \ce{Si} atoms are primarily bonded to both an \ce{OH} and a \ce{CH3} group. 
In contrast, for paired (clustered) distributions, a positive mixing enthalpy is observed. 
Moreover, as the cluster size increases (marked in yellow), the mixing enthalpy gradually decreases, approaching the equilibrium line, suggesting that phase separation becomes more favorable. 
This confirms that an unpaired functionalization pattern is thermodynamically favored at finite temperature. 

The obtained distribution of mixing enthalpies closely resembles the experimental trend in surface free energy, as shown in Figure~\ref{fig:mixing_energy}, measured for \ce{OH}/\ce{CH3}-functionalized silica surfaces by Yang et al.~\cite{Yang2009}. 
In the experimental data, a similar negative deviation from ideal mixing is observed, with a minimum at approximately 57\% \ce{CH3} surface coverage. 
This value corresponds to the composition at which the measured surface free energy is lowest compared to the linear combination of the pure \ce{OH}- and \ce{CH3}-terminated surfaces. 
The surface mixing enthalpies at intermediate compositions were computed as the difference between the actual surface free energy and the value predicted for an ideal mixture of the pure phases. 
Although this effect was not discussed in the original study, we interpret it as an indication of a comparable stabilization trend for mixed functionalizations. 
Despite the difference in scale, this agreement supports the conclusion that unpaired distributions of \ce{OH} and \ce{CH3} groups are thermodynamically favored at intermediate compositions.
However, further analysis of the underlying mechanism is necessary, as the stabilization of this specific pattern is likely driven by the formation of a hydrogen-bond network at the surface. 

\subsection{Hydrogen Bonding Network}
\label{subsec5}

\begin{figure}[!htb]
\centering
     \includegraphics[width=0.8\textwidth]{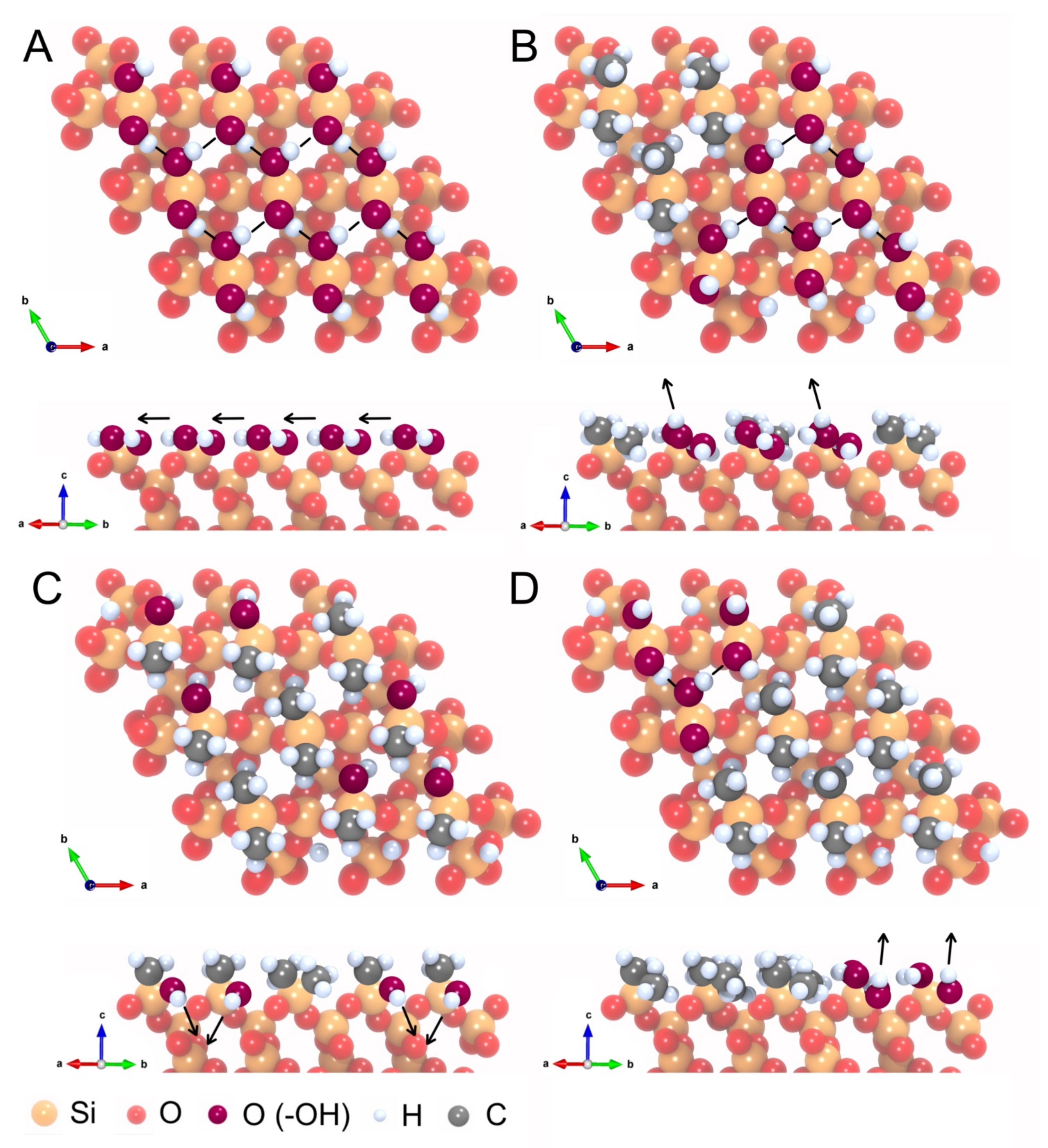}
\caption{\label{fig:H-bonds} 
Snapshots of the hydrogen bonding arrangement for the functionalized $\alpha$-\ce{SiO2} (0001) surface with varying \ce{OH}/(\ce{OH}+\ce{CH3}) ratios: A - $100$ \% \ce{OH}, B - $67$ \% \ce{OH} (paired, cluster), C - $33$ \% \ce{OH} (unpaired) and D - $33$ \% \ce{OH} (paired). The silicon atoms are shown in yellow, framework oxygen atoms (\ce{SiO2}) in orange, hydroxyl oxygen atoms in red, hydrogen atoms in white, and carbon atoms in grey. The arrows indicate the direction of \ce{OH} groups forming the hydrogen bond network. The solid black line denotes the characteristic zig-zag signature of the hydrogen‐bonding network on the fully hydroxylated surface.
}
\end{figure}

\begin{figure}[!htb]
\centering
     \includegraphics[width=0.75\textwidth]{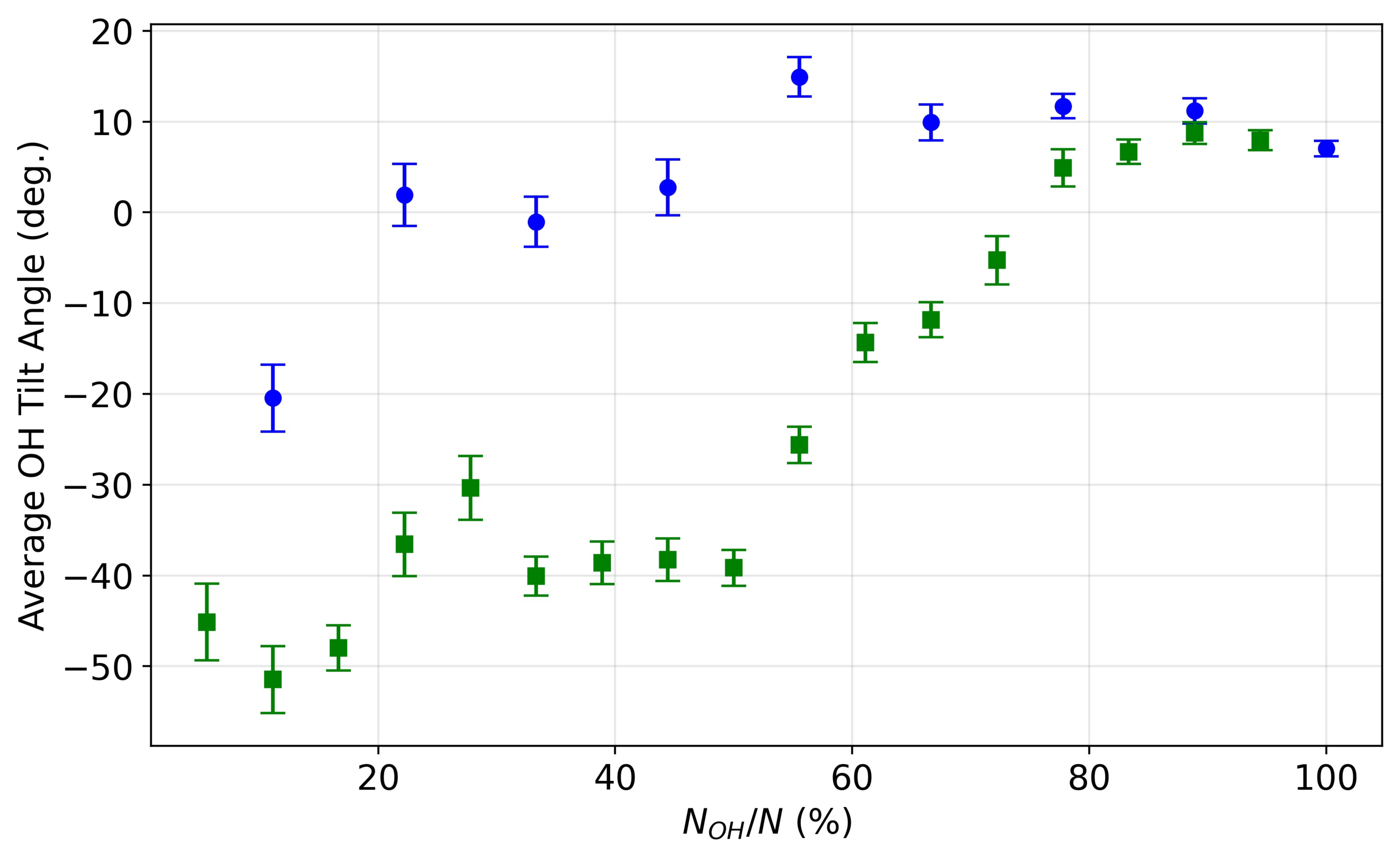}
\caption{\label{fig:tilt_analysis} 
Average tilt of \ce{OH} groups depending on the  \ce{OH}/\ce{CH3} functional group distributions on the $\alpha$-\ce{SiO2} (0001) surface, calculated from MD simulations at $300$ K. Each data point represents the average over $2$ independent surface models, with statistics collected from $80000$ MD steps. Green colour represents unpaired \ce{OH-Si-CH3} configurations, while blue indicates paired \ce{Si-(OH)2}/\ce{Si-(CH3)2} configurations.}
\end{figure}

The structural arrangement and geometric features of the hydrogen bonding network are depicted in Figure~\ref{fig:H-bonds} and summarized in Table~\ref{tab:hh_dist}, focusing on key \ce{H-O \cdots H} interactions. 
The average tilt of \ce{OH} groups as a function of the  \ce{OH}/\ce{CH3} functional group distributions is illustrated in Figure \ref{fig:tilt_analysis}. 
To quantitatively assess the hydrogen‑bond parameters, we developed a Python script, which is openly available on GitHub.\cite{strugovshchikov2025hydrogen}

\begin{table}[!htb]
\renewcommand{\arraystretch}{1.2}
\tabcolsep=0.16cm  
\centering
\caption{\label{tab:hh_dist} Hydrogen bonding network analysis in terms of \ce{O-H}$\cdot \cdot \cdot$\ce{O} connectivities, represented by the average hydrogen bond distance (\ce{O-H}$\cdot \cdot \cdot$\ce{O}) and angle ($\angle,$\ce{O-H}$\cdot \cdot \cdot$\ce{O}) across all productive MD simulation steps. The \ce{OH} tilt is defined as the average angle between the \ce{OH} groups and the \textit{xy} plane of the surface. Each data point represents the average over $2$ independent surface models, with statistics collected from $80000$ MD steps.}
\footnotesize  
\begin{tabular}{c c c c c c c} 
\toprule
\ce{N_{OH}}/N & \multicolumn{2}{c}{\ce{OH} tilt ($^{\circ}$)} & \multicolumn{2}{c}{\ce{O-H}$\cdot \cdot \cdot$\ce{O} (\AA)} & \multicolumn{2}{c}{$\angle\,$\ce{O-H}$\cdot \cdot \cdot$\ce{O} ($^{\circ}$)} \\
\cmidrule(lr){2-3} \cmidrule(lr){4-5} \cmidrule(lr){6-7}
(\%) & Paired & Unpaired & Paired & Unpaired & Paired & Unpaired \\
\midrule
$100$ & $7.0 \pm 0.8$ & $-$ & $1.99 \pm 0.03$ & $-$ & $159.8 \pm 1.1$ & $-$ \\
$94$ & $-$ & $8.0 \pm 1.1$ & $-$ & $1.91 \pm 0.03$ & $-$ & $160.2 \pm 1.1$ \\
$89$ & $11.2 \pm 1.4$ & $8.7 \pm 1.2$ & $1.91 \pm 0.02$ & $1.87 \pm 0.02$ & $158.7 \pm 1.2$ & $160.6 \pm 1.0$ \\
$83$ & $-$ & $6.7 \pm 1.4$ & $-$ & $1.92 \pm 0.02$ & $-$ & $159.5 \pm 1.1$ \\
$78$ & $11.7 \pm 1.3$ & $4.9 \pm 2.0$ & $1.90 \pm 0.02$ & $1.87 \pm 0.04$ & $159.2 \pm 1.2$ & $158.9 \pm 1.2$ \\
$72$ & $-$ & $-5.3 \pm 2.7$ & $-$ & $1.96 \pm 0.02$ & $-$ & $156.7 \pm 1.3$ \\
$67$ & $9.9 \pm 2.0$ & $-11.8 \pm 1.9$ & $1.85 \pm 0.03$ & $1.96 \pm 0.02$ & $159.6 \pm 1.3$ & $157.2 \pm 1.3$ \\
$61$ & $-$ & $-14.3 \pm 2.1$ & $-$ & $1.98 \pm 0.02$ & $-$ & $156.9 \pm 1.3$ \\
$56$ & $14.9 \pm 2.2$ & $-25.6 \pm 2.0$ & $1.98 \pm 0.03$ & $1.99 \pm 0.02$ & $154.8 \pm 1.9$ & $157.0 \pm 1.4$ \\
$50$ & $-$ & $-39.2 \pm 2.0$ & $-$ & $2.02 \pm 0.02$ & $-$ & $155.2 \pm 1.5$ \\
$44$ & $2.8 \pm 3.1$ & $-38.3 \pm 2.3$ & $2.00 \pm 0.04$ & $1.99 \pm 0.02$ & $156.2 \pm 1.8$ & $155.5 \pm 1.6$ \\
$39$ & $-$ & $-38.6 \pm 2.4$ & $-$ & $1.97 \pm 0.02$ & $-$ & $155.7 \pm 1.7$ \\
$33$ & $-1.0 \pm 2.8$ & $-40.1 \pm 2.2$ & $2.06 \pm 0.04$ & $1.95 \pm 0.03$ & $153.1 \pm 2.3$ & $155.7 \pm 1.8$ \\
$28$ & $-$ & $-30.3 \pm 3.5$ & $-$ & $1.93 \pm 0.03$ & $-$ & $154.8 \pm 2.1$ \\
$22$ & $1.9 \pm 3.4$ & $-36.6 \pm 3.5$ & $1.90 \pm 0.06$ & $1.94 \pm 0.04$ & $157.5 \pm 2.9$ & $153.8 \pm 2.3$ \\
$17$ & $-$ & $-48.0 \pm 2.5$ & $-$ & $1.81 \pm 0.03$ & $-$ & $156.8 \pm 2.6$ \\
$11$ & $-20.5 \pm 3.7$ & $-51.5 \pm 3.7$ & $2.07 \pm 0.08$ & $1.85 \pm 0.04$ & $154.0 \pm 3.3$ & $154.1 \pm 3.3$ \\
$6$ & $-$ & $-45.1 \pm 4.2$ & $-$ & $1.85 \pm 0.06$ & $-$ & $155.7 \pm 4.6$ \\
\bottomrule
\end{tabular}
\end{table}

First, we analyze the hydrogen bond network of the silica surface fully covered with \ce{OH} groups. 
The results indicate that the formation of an interconnected zigzag-like \ce{OH} network facilitates non-covalent interactions between hydroxyl protons and the hydroxyl oxygens, leading to a robust hydrogen-bonding network (Figure~\ref{fig:H-bonds}A). 
According to Figure~\ref{fig:tilt_analysis}, the \ce{OH} groups are predominantly aligned along the surface \textit{xy}-plane, with an average tilt angle of $7.0^{\circ} \pm 0.8^{\circ}$ at $300$ K. 
According to previously published results\cite{Yang2006}, the hydrogen bonds can be classified into two types based on their strength: a stronger type with an \ce{O}-\ce{H} distance of $1.79$ \AA~ and an angle ($\angle,$\ce{O-H}$\cdot \cdot \cdot$\ce{O}) of $167^{\circ}$, and the weaker type with \ce{O}-\ce{H} distance of $2.19$ \AA~ and an angle ($\angle,$\ce{O-H}$\cdot \cdot \cdot$\ce{O}) of $165^{\circ}$. 
As is apparent, the average hydrogen bond lengths and angles gathered in Table~\ref{tab:hh_dist} closely match previously reported values, thus confirming the reliability of our structural model as a starting point for analyzing the effect of an increasing \ce{CH3} content in mixed functionalizations.

The introduction of \ce{CH3} groups is expected to significantly alter the hydrogen bonding interactions of \ce{OH} groups, with distinct effects depending on the functional group distribution. 
At low \ce{CH3} concentrations, the tilt angle of \ce{OH} groups remains largely unchanged, indicating that the hydrogen-bonding network is mostly preserved. 

On the other hand, at a \ce{CH3} content of $28$ \% ($72$ \% of \ce{OH}), a notable difference emerges between the paired and unpaired distributions of \ce{OH} groups. 
In the unpaired configuration, the average tilt of \ce{OH} groups engaged into hydrogen bonds decreases significantly, accompanied by a slight increase in the average hydrogen bond length. 
This result suggests that unpaired \ce{OH} groups tend to reorient toward subsurface oxygen atoms, forming a secondary hydrogen-bonding network with the silica sublayer.

In contrast, for the paired configuration, the insertion of \ce{CH3} groups does not lead to significant reorientation of the \ce{OH} groups. 
Instead, the \ce{OH} groups predominantly lose their hydrogen bonds and remain tilted away from the surface,  rather than reorienting toward subsurface oxygen atoms to form new interactions. 
Reorientation becomes evident only at higher \ce{CH3} concentrations, when the remaining hydroxyl groups exist as isolated \ce{Si-(OH)2} pairs fully encased by methyl \ce{CH3} groups.

These results reveal two distinct hydrogen‐bonding modes dictated by local functionalization patterns. 
In paired configurations, which are energetically favored when \ce{OH} groups cluster, the silanol network remains internally stabilized: \ce{OH} groups within the cluster preserve their hydrogen bonds, whereas those at the cluster–\ce{CH3} boundary lose \ce{H}‐bonding and tilt away from the surface without finding new donors. 
In contrast, in unpaired distributions, isolated \ce{OH}-\ce{Si}-\ce{CH3} sites permit the remaining \ce{OH} groups to reorient and form hydrogen bonds with subsurface oxygen atoms. 
This behavior arises because the lower local \ce{OH} concentration in unpaired regions prevents intra‐cluster bonding, and adjacent \ce{CH3} groups sterically hinder lateral \ce{H}‐bond formation, effectively steering the \ce{OH} protons toward subsurface oxygens.

\subsection{Vibrational spectra analysis}
\label{subsec6}

\begin{figure}[!htb]
\centering
     \includegraphics[width=0.85\textwidth]{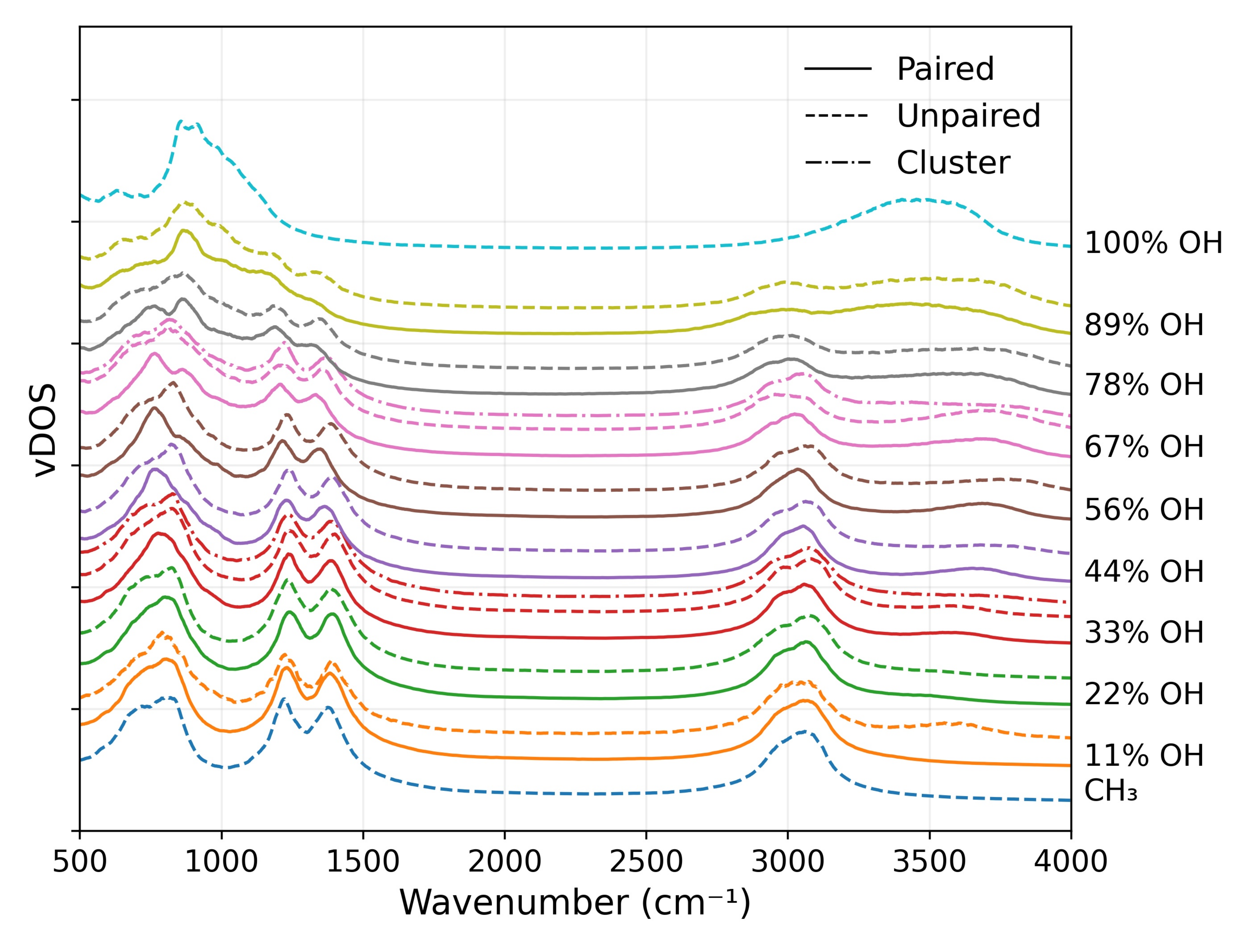}
\caption{\label{fig:vibrational} 
Vibrational density of states (vDOS) computed for the functional groups on $\alpha$-\ce{SiO2} surfaces functionalized with varying \ce{OH}/\ce{CH3} ratios.
Solid lines represent unpaired configurations, dashed lines indicate paired configurations, and dash-dotted lines correspond to clustered configurations. }
\end{figure}

To further assess the character of hydrogen bonding (HB) network in the functionalized \ce{SiO2}, we analyzed the vibrational frequencies of the functional groups.
Figure~\ref{fig:vibrational} presents the vibrational density of states (vDOS) for functionalized surfaces with varying \ce{OH}/\ce{CH3} ratios.
The vibrational spectra were obtained by performing Fourier transforms of the time-dependent bond and angle fluctuations extracted from the MD trajectories of the model system\cite{Ditler2022}.
This approach enables a flexible decomposition of the total vibrational spectrum into contributions from different functional groups. 
To compute the vibrational spectra, we developed a Bash/Python script, which is available on GitHub\cite{strugovshchikov2025vibrational}.
In particular, we focus on the $3200$–$3800$ cm$^{-1}$ region of the spectrum, corresponding to the \ce{OH} stretching modes, where frequency shifts can serve as sensitive indicators of hydrogen bond strength and configuration.

\begin{figure}[!htb]
\centering
     \includegraphics[width=0.85\textwidth]{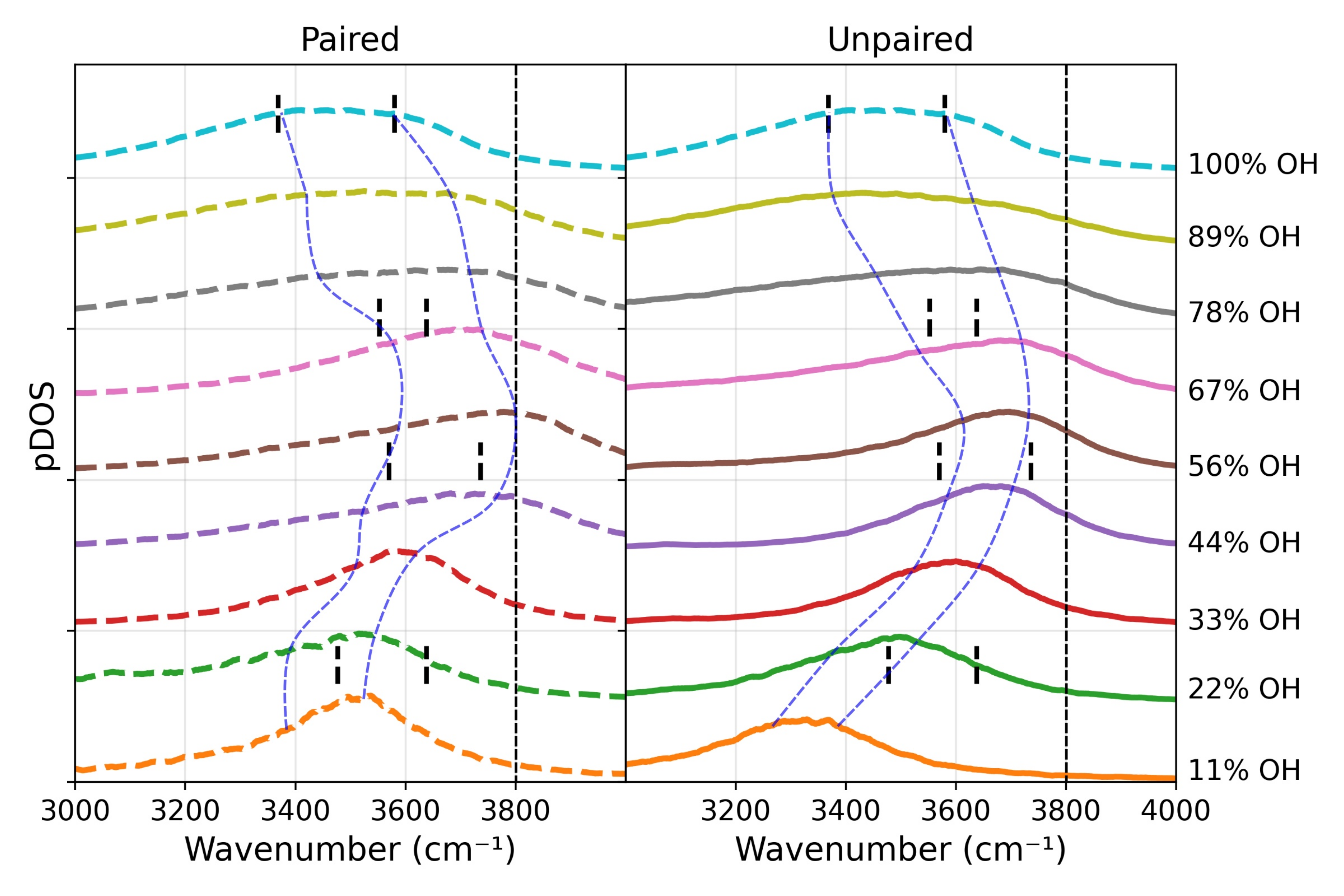}
\caption{\label{fig:vibrational_OH} 
Partial vibrational density of states (pDOS) for \ce{OH} contributions on $\alpha$-\ce{SiO2} surfaces functionalized with varying \ce{OH}/\ce{CH3} ratios. Dashed lines correspond to paired configurations, while solid lines represent unpaired arrangements. Blue dashed markers indicate the characteristic peak positions of theoretically computed \ce{OH} stretching modes, capturing the influence of local hydrogen-bonding environments. Black vertical dashed lines denote experimental peak positions adapted from Yang~\cite{Yang2009}, shifted upward by 140~cm$^{-1}$. In particular, the rightmost black dashed line highlights the experimental signature of non-bonded residual \ce{OH} groups, indicative of weakened or absent hydrogen bonding.}
\end{figure}

First, the effect of the hydrogen bond network on vibrational properties was demonstrated by analyzing the fully hydroxylated surface. 
A critical feature of hydrogen bonding is the redshift in the stretching frequencies of proton-donating \ce{OH} groups relative to free (non-hydrogen bonded) \ce{OH} groups. 
To isolate the spectral contributions of \ce{OH} groups, we also computed the partial vibrational density of states (pDOS) corresponding to the isolated \ce{O}-\ce{H} stretching mode, as shown in Figure~\ref{fig:vibrational_OH}. 
For the fully hydroxylated surface, the vibrational spectrum reveals a set of high-frequency modes corresponding to decoupled \ce{OH} stretching vibrations, with observed peaks at $3383$ and $3578$ cm$^{-1}$. 
These values align well with experimental observations reported by Yang et al.,\cite{Yang2009} who used FT-IR spectroscopy to study the influence of the \ce{OH}/\ce{CH3} ratio on the vibrational properties of methyl-modified silica films.
In their study, two distinct \ce{OH}-stretching modes were identified at approximately $3228$ and $3440$ cm$^{-1}$, attributed to hydrogen bonds of varying strength. 
To account for known systematic errors in DFT, particularly the overestimation of vibrational frequencies due to the neglect of nuclear quantum effects, red shifts on the order of $100$–$150$ cm$^{-1}$ are commonly applied to align calculated frequencies with experiment. 
For example, Ojha et al. observed a red shift of 120 cm$^{-1}$ in OH-stretch modes when including quantum nuclear motion via path-integral MD simulations \cite{Ojha2024}. 
After applying the $140$ cm$^{-1}$ correction, the relative separation between the two modes remains consistent with the computed spectra, indicating that the variation in hydrogen bonding environments is well captured by the simulations.
These results also indicate a redshift of approximately $300$ cm$^{-1}$ compared to the stretching mode of an isolated hydroxyl group on \ce{SiO2}\cite{Morrow1973}. 
The experimental signature of non-bonded residual \ce{OH} groups, corresponding to this mode, is shown in Figure~\ref{fig:vibrational_OH} as a black dashed line. 
This shift further supports the presence of a hydrogen-bonding network, in which protonic hydrogen atoms interact with neighboring oxygen atoms from adjacent \ce{OH} groups. 
Thus, the computed vibrational frequencies for the fully hydroxylated surface provide strong evidence for the formation of an extended hydrogen bond network.

Extending this analysis to partially functionalized surfaces, we examined the vibrational behavior of \ce{OH} groups in systems with mixed \ce{OH}/\ce{CH3} distributions.
The introduction of \ce{CH3} groups into the system leads to the emergence of additional peaks in the vibrational spectra, which become more pronounced with increasing \ce{CH3} content. 
Specifically, characteristic peaks appear at approximately $1250$ cm$^{-1}$ and $1410$ cm$^{-1}$, corresponding to \ce{Si-CH3} bending modes, as well as at $3000$ cm$^{-1}$ and $3100$ cm$^{-1}$, attributed to \ce{C-H} stretching vibrations. 
The FT-IR measurements of methyl-modified silica films revealed absorption peaks at $1276$ and $1412$ cm$^{-1}$ associated with the \ce{Si-CH3} band, and a peak at $2978$ cm$^{-1}$ corresponding to \ce{C-H} vibrations of methyl groups \cite{Yang2009}. 
Their results show that increasing the \ce{OH}/\ce{CH3} ratio leads to an enhancement of the intensity of the \ce{Si-CH3} absorption peaks, while simultaneously weakening the \ce{O-H} absorption bands.  

Consistent with experimental observations, our simulations reveal that increasing \ce{CH3} substitution alters the hydrogen‐bonding environment of surface \ce{OH} groups, as evidenced by blue shifts in the \ce{O}-\ce{H} stretching frequencies relative to the fully hydroxylated surface. 
Figure~\ref{fig:vibrational_OH} compares the partial vibrational spectra for \ce{OH} contributions in paired and unpaired configurations.
In the unpaired configuration, the \ce{O}–\ce{H} stretch exhibits a maximum at $3708$ cm$^{-1}$, indicative of weakened but persistent hydrogen bonding directed toward subsurface oxygen atoms. 
In contrast, the paired configuration shows a stronger blue shift to 3797 cm$^{-1}$, corresponding to a complete loss of intra-cluster hydrogen bonding and a reorientation of the \ce{OH} proton away from any hydrogen-bond acceptor. 

These differences in spectral response reflect the varying degrees of hydrogen-bond disruption and local structural flexibility. 
The unpaired configuration retains some residual hydrogen bonding and exhibits a distinct tilt of \ce{OH} groups, whereas the more rigid paired configuration lacks such interactions. 
Notably, Yang et al.\cite{Yang2009} reported similar blue shifts in the experimental FT-IR spectra with increasing \ce{CH3} content, with peak positions progressing from $3440$ cm$^{-1}$ in hydroxyl-rich surfaces to 3633 cm$^{-1}$. 
The progressive frequency increase observed in our simulations closely mirrors this experimental trend as illustrated in Figure~\ref{fig:vibrational_OH}.

Importantly, the unpaired configurations in our model show better agreement with the experimental peak positions, underscoring the role of configurational diversity in capturing the vibrational response. 
These findings highlight a limitation of conventional surface models, which typically assume only idealized paired functional groups and therefore fail to capture key effects arising from local disorder and surface flexibility. 
In contrast, our machine-learned force field provides a more realistic description of the mixed \ce{OH}/\ce{CH3}-functionalized silica surface, leading to improved consistency with experimental observations.
By revealing how the spatial patterning of surface groups modulates hydrogen bonding and vibrational signatures, our study advances the fundamental understanding of interfacial interactions at chemically heterogeneous surfaces, which is a central focus in interfacial processes, capillarity, and wetting phenomena.

\section*{Conclusion}

In this study, we investigated the atomic-scale functionalization of $\alpha$-quartz (\ce{SiO2}) (0001) surfaces with hydroxyl (\ce{OH}) and methyl (\ce{CH3}) groups, focusing on the role of their spatial distribution.  
We show that not only the concentration but also the arrangement of functional groups significantly impacts surface thermodynamics, hydrogen bonding, and vibrational signatures.  
A particularly favorable unpaired \ce{OH}/\ce{CH3} configuration emerges near 67\% \ce{CH3} substitution, where reoriented unpaired \ce{OH} groups form secondary hydrogen bonds.  
This is reflected in distinct blue shifts in vibrational \ce{OH} stretching modes, consistent with weaker hydrogen bonding.

Our model introduces a conceptual advance by explicitly capturing spatial heterogeneity in surface functionalization, an aspect often overlooked in classical force fields and previous theoretical models\cite{Khanniche2017, Cygan2004,Abramov2019, Furukawa2005}.  
For instance, the force field of Khanniche et al.~\cite{Khanniche2017} lacks parameters for mixed \ce{OH}–\ce{Si}–\ce{CH3} motifs, making it unable to reproduce the cooperative effects revealed here.
Our MLFF-based approach enables quantum-accurate simulations of chemically complex silica surfaces and provides a predictive framework for tuning interfacial energy and wettability.  
To our knowledge, this is the first demonstration of a machine-learned potential capable of capturing both thermodynamic and vibrational trends across a wide range of \ce{OH}/\ce{CH3} compositions.  
The computed mixing energies (Figure~\ref{fig:mixing_energy}) align closely with experimental surface free energy data~\cite{Yang2009}, and the vibrational spectra (Figure~\ref{fig:vibrational_OH}) reproduce experimentally observed frequency shifts, validating our treatment of local hydrogen-bonding environments.  
Together, these results establish a quantitative foundation for tailoring hybrid silica interfaces through local patterning.

Ongoing work investigates how these atomic-scale arrangements influence water contact angles and interfacial structuring to bridge molecular design with experimental observables.  
We also plan to incorporate water–surface interactions and adsorption dynamics into MLFF training, enabling more accurate simulations under realistic aqueous or gaseous conditions.  
Future studies will explore how varying \ce{OH}/\ce{CH3} ratios and hydrogen-bonding patterns affect surface reconstruction~\cite{Malyi2014,Murashov2005}, helping to clarify the interplay between chemical functionalization and structural evolution.  
Altogether, our findings provide new design principles for tuning colloidal interactions, wetting behavior, and adhesion on silica-based materials, of direct interest to the colloid and interface science community.


\section*{Acknowledgements}
\label{app1}

This research is supported by ANR project “PROMENADE” No. ANR-23-CE50-0008 and by the ANR French PIA project “Lorraine Université d’Excellence” No. ANR-15-IDEX-04-LUE. 
Molecular simulations were conducted using HPC resources from GENCI-TGCC and GENCI- IDRIS (eDARI projects No. A0150913052 and A0170913052), as well as resources provided by the EXPLOR Center hosted by the University of Lorraine.

\appendix
\section{Associated content}
\label{app2}

\subsection*{Supporting Information}

The Supporting Information is available free of charge. 
It includes the details and results of the DFT calculations used for geometry optimizations.

\subsection*{Data Availability}

The training dataset and machine learning force field developed in this study are openly available at: https://doi.org/10.5281/zenodo.16883356.

The training set written according to VASP internal standards (ML\_AB file), including geometry data, energies, forces, and stresses.  
The machine learning force field (ML\_FF).

\subsection*{Code Availability}

All scripts are openly available on GitHub:
\begin{itemize}
  \item Vibrational spectra of OH and CH$_3$ on $\alpha$‑\ce{SiO2}
  
  (\texttt{Vibrational\_OH\_CH3\_SiO2\_VASP}),
  \item Hydrogen‑bond network parameters of OH on $\alpha$‑\ce{SiO2}
  
  (\texttt{Hydrogen\_bond\_network\_OH\_SiO2}).
\end{itemize}

\section*{Author Contributions}
\label{app3}

The manuscript was written through contributions of all authors. 
All authors have given approval to the final version of the manuscript.

\section*{Notes}
\label{app4}

The authors declare no competing financial interest.

\bibliographystyle{elsarticle-num} 
\bibliography{main}         

\end{document}